\documentstyle[aps,twocolumn]{revtex}

\begin{document}
\title{Calculation of entanglement for continious variable states}
\author{Huai-Xin Lu,$^{1,2}$ Zeng-Bing Chen,$^1$ Jian-Wei Pan,$^{1,3}$ and Yong-De
Zhang$^{4,5}$}
\address{$^1$Department of Modern Physics, University of Science and Technology of
China, Hefei, Anhui 230027, China $^2$Department of Physics ,Wefang
University, Wefang, Shandong 261043, China\\
$^3$Institut f\"ur Experimentalphysik, Universit\"at Wien, Boltzmanngasse 5,
1090 Wien, Austria\\
$^4$CCAST (World Laboratory), P.O. Box 8730, Beijing 100080, China \\
$^5$Laboratory of Quantum Communication and Quantum Computation and
Department of Modern Physics, University of Science and Technology of China,
Hefei, Anhui 230027, China}
\date{}
\maketitle

\begin{abstract}
In this paper, we present a general formula for obtaining the reduced
density opeator for any biparticle pure entangled state. Using this formula,
we derive, in a compact form, the explicit formula of the entanglement for
any bipartical pure entangled Gaussian state. In the case of Gaussian
states, the criteria of separabelity can be naturely obtained by the
formula. For non-Gaussian states, we also show the usefulness of the method
presented in this paper.
\end{abstract}

Quantum entanglement is one of the essential features of quantum mechanics.
It has been an interesting and important topic since it was first noted by
Einstein, Podolsky, and Rosen (EPR) \cite{EPR}. Now, it has attracted much
attention because of immense progresses both in the foundation of quantum
mechanics and in the burgeoning field of quantum information theory. Issues
such as the relationship between entanglement and quantum nonlocality and
the violations of Bell's inequality for both discrete \cite{Bell,CHSH} and
continuous variable states \cite{Banaszek} further enrich the contents of
quantum mechanics. In recent years, quantum entanglement is also viewed as a
useful ``resource'' for various kinds of quantum information processing. The
original EPR states of continuous variables can be created either by the
nondegenerate optical parametric amplifier \cite{Reid,Ou,Ou-APB}, or simply
by using passive optical elements (e.g., beam splitters) \cite
{CV-network,CV-BS,Kim}. While most quantum information protocols were
initially developed for quantum systems with discrete quantum variables,
quantum information processing based on quantum states with continuous
variable has also been proposed, e.g., quantum teleportation \cite{BK},
quantum error correction \cite{error}, quantum computation \cite{computation}%
, entanglement purification \cite{Duan} and cloning \cite{cloning}. Thus,
inseparability criteria \cite{Duan-c,Simon,WangXB} and entanglement
quantification \cite{Parker} in continuous variable systems has become an
issue of practical importance.

The measurement of entanglement for any biparticle pure entangled state,
i.e., the von Neumann entropy of either partial trace of the density
operator for the state, is considered as a ``good measurement'' of quantum
entanglement. Parker {\it et al}. \cite{Parker} have developed an elegant
method to calculate the entanglement of the biparticle pure entangled states
with continuous variables by means of the integral eigenvalue equations in
the coordinate-momentum space. In fact, their method is based on the Schmidt
decompositions. But in certain cases, it is somewhat harder to find such a
``Schmidt basis'' for the continuous variable systems. In this paper, we
propose an alternative method to calculate the entanglement entropy for
continuous variable states in Fock space. With the help of linear quantum
transformation theory (LQTT) \cite{zhang}, we obtain an explicit formula of
the entanglement entropy for {\it any} Gaussian state. With this formula, we
can easily calculate the Gaussian states' entanglement without knowing the
eigenvalues of the reduced density operator for the system. Furthermore,
from the derivation of the formula, we naturally obtain the necessary and
sufficient condition of separability for a biparticle pure entangled state
with continuous variables. For some non-Gaussian states, one can also
calculate the eigenvalues of the reduced density operators, from which the
entanglement of this bipartite system may be easily obtained.

It is well known that the entanglement entropy of a biparticle pure
entangled state is defined by 
\begin{equation}
E=-%
\mathop{\rm tr}
\rho _1\ln \rho _1=-%
\mathop{\rm tr}
\rho _2\ln \rho _2,  \label{entangled}
\end{equation}
where the reduced density operator $\rho _{1(2)}$ is a partial trace of
density operator $\rho _{12}$ for the state, i.e., 
\begin{equation}
\rho _{1(2)}=%
\mathop{\rm tr}
\nolimits_{2(1)}\rho _{12}.  \label{trace}
\end{equation}
In order to give a general formula of reduced density operator $\rho _1$ for
any biparticle system, we use the following formula \cite{zhang,Zuber} 
\begin{eqnarray}
\Omega &=&:\exp \left[ (a_1^{\dagger },a_2^{\dagger })%
{\partial _{z_1^{*}} \choose \partial _{z_2^{*}}}
+\left( \partial _{z_1},\partial _{z_2}\right) 
{a_1 \choose a_2}
\right] :  \nonumber  \label{addition1} \\
&&\ \left. \times \left\langle Z\left| \Omega \right| Z\right\rangle \mid
_{z=0},\right.  \label{addition1}
\end{eqnarray}
where $\Omega $ is an arbitrary bosonic operator, $a_i^{\dagger }$ $(i=1,2)$
is the usual bosonic creation operator for the subsystem $i$; $\left|
Z\right\rangle =\left| Z_1\right\rangle \otimes \left| Z_2\right\rangle $
denote the usual coherent states, $:\;:$ means the normal ordering and $z=0$
means $z_{1(2)}=z_{1(2)}^{*}=0$. Using Eq. (\ref{addition1}), we write the
density operator $\rho _{12}$ for the biparticle system as 
\begin{eqnarray}
\rho _{12} &=&:\exp \left[ (a_1^{\dagger },a_2^{\dagger })%
{\partial _{z_1^{*}} \choose \partial _{z_2^{*}}}
+\left( \partial _{z_1},\partial _{z_2}\right) 
{a_1 \choose a_2}
\right] :  \nonumber  \label{addition2} \\
&&\ \left. \times \left\langle Z\left| \rho _{12}\right| Z\right\rangle \mid
_{z=0}\right. ,  \label{addition2}
\end{eqnarray}
where $\left| Z\right\rangle \equiv \left| Z_1,Z_{2\text{ }}\right\rangle $.
With the help of overcomplete of coherent states $\int \frac{d^2Z_2}\pi
\left| Z_2\right\rangle _{22}\left\langle Z_2\right| =1$, and taking the
partial trace over subspace of the subsystem $2$, one can get the reduced
density operator $\rho _1$ as follows 
\begin{eqnarray}
\rho _1 &=&%
\mathop{\rm tr}
\nolimits_2\rho _{12}=%
\mathop{\rm tr}
\nolimits_2\int \frac{d^2z_2^{^{\prime }}}\pi \left| Z^{^{\prime
}}\right\rangle _{22}\left\langle Z^{^{\prime }}\right|  \nonumber \\
&&\left. \times :e^{(a_1^{\dagger },a_2^{\dagger })%
{\partial _{z_1^{*}} \choose \partial _{z_2^{*}}}
+\left( \partial _{z_1},\partial _{z_2}\right) 
{a_1 \choose a_2}
}:\left\langle Z\left| \rho _{12}\right| Z\right\rangle \mid _{z=0}\right. 
\nonumber \\
\ &=&\int \frac{d^2z_2^{^{\prime }}}\pi :\exp \left( a_1^{\dagger }\partial
_{z_1^{*}}+\partial _{z_1}a_1\right) :  \nonumber \\
&&\left. \times \exp \left( z_2^{^{\prime }*}\partial _{z_2^{*}}+\partial
_{z_2}z_2^{^{\prime }*}\right) \left\langle Z\left| \rho _{12}\right|
Z\right\rangle \mid _{z=0}\right. .  \label{addition3}
\end{eqnarray}

In the following we will consider two classes of states: the Gaussian states
and the non-Gaussian ones. Let us first consider the following Gaussian
states 
\begin{eqnarray}
\rho _{12} &=&A_0:\exp \left\{ \frac 12\left[ (a_1^{\dagger },a_1)M_1%
{a_1^{\dagger } \choose a_1}
\right. \right.  \nonumber  \label{density} \\
&&\left. \left. \left. +(a_2^{\dagger },a_2)M_2%
{a_2^{\dagger } \choose a_2}
+2(a_1^{\dagger },a_1)M_{12}%
{a_2^{\dagger } \choose a_2}
\right] \right\} :\right. ,  \label{density}
\end{eqnarray}
where $A_0$ is a normalization factor, $M_i$ $(i=1,2)$ is Hermitian
matrices, and 
\[
M_{12}=\left( 
\begin{tabular}{ll}
$e$ & $f$ \\ 
$f^{*}$ & $e^{*}$%
\end{tabular}
\right) 
\]
with $e$ and $f$ being two arbitrary complex numbers. In Eq. (\ref{density})
the linear terms of $a$ and $a^{\dagger }$ are not included since they do
not affect the entanglement of the biparticle Gaussian states \cite{Duan-c}.
In this case, the matrix elements of $\rho _{12}$ with respect to $\left|
Z\right\rangle $ is 
\begin{eqnarray}
\left\langle Z\left| \rho _{12}\right| Z\right\rangle &=&A_0\exp \left\{ 
\frac 12\left[ (z_1^{*},z_1)M_1%
{z_1^{*} \choose z_1}
+(z_2^{*},z_2)\right. \right.  \nonumber  \label{addition4} \\
&&\left. \left. \times M_2%
{z_2^{*} \choose z_2}
+2(z_1^{*},z_1)M_{12}%
{z_2^{*} \choose z_2}
\right] \right\} .  \label{addition4}
\end{eqnarray}
Using the following Gaussian integration formula \cite{pan} 
\begin{eqnarray}
&&\int \frac{d^2z}\pi \exp \left\{ -\frac 12(z^{*},z)Q%
{z^{*} \choose z}
+(u,v)%
{z^{*} \choose z}
\right\}  \nonumber \\
\ &=&\left[ -\det Q\right] ^{\frac{-1}2}\exp \left\{ \frac 12(u,v)Q^{-1}%
{u \choose v}
\right\} ,  \label{integerate}
\end{eqnarray}
where $Q=\widetilde{Q}$ is nonsingular and $u$ ($v$) is an arbitrary complex
number, and the formula 
\begin{eqnarray}
&&\left. \exp \left( z_2^{^{\prime }*}\partial _{z_2^{*}}+\partial
_{z_2}z_2^{^{\prime }}\right) \Psi (z_1^{*},z_1;z_2^{*},z_2)\mid
_{z_2=z_2^{*}=0}\right.  \nonumber  \label{displace} \\
\ &=&\left. \Psi (z_1^{*},z_1;z_2^{^{\prime }*},z_2^{^{\prime }})\right. ,
\label{partialzz}
\end{eqnarray}
we have, after substituting Eq. (\ref{addition4}) into Eq. (\ref{addition3}%
), 
\begin{equation}
\rho _1=\frac{A_0}{\sqrt{-\det M_2}}:e^{\frac 12(a_1^{\dagger
},a_1)(M_1-M_{12}M_2^{-1}\tilde M_{12})%
{a_1^{\dagger } \choose a_1}
}:.  \label{partial2}
\end{equation}
According to LQTT \cite{zhang}, if one denotes 
\begin{equation}
M=\left( 
\begin{tabular}{ll}
$a$ & $d$ \\ 
$b$ & $c$%
\end{tabular}
\right) ,  \label{matrix1}
\end{equation}
there is a map of $M$ defined as 
\begin{eqnarray}
D(M) &=&(M_1-M_{12}M_2^{-1}\tilde M_{12})\Sigma _B^{-1}  \nonumber
\label{dom} \\
\ &=&\left( 
\begin{tabular}{ll}
$c^{-1}-1$ & $c^{-1}d$ \\ 
$c^{-1}b$ & $1-c^{-1}$%
\end{tabular}
\right) \Sigma _B^{-1},  \label{dom}
\end{eqnarray}
where complex constants $a,b,c$ and $d$ are determined by the known matrices 
$M_{1(2)}$ and $M_{12}$ in Eq. (\ref{partial2}). Then we can immediately
rewrite the normally ordered form in Eq. (\ref{partial2}) as \cite{zhang} 
\begin{equation}
\rho _1=A\exp \left[ \frac 12(a_1^{\dagger },a_1)N\Sigma _B%
{a_1^{\dagger } \choose a_1}
\right] ,  \label{partial3}
\end{equation}
where negative Hermitian matrix $N=\ln M$, $\sum_B=\left( 
\begin{tabular}{ll}
$0$ & $1$ \\ 
$-1$ & $0$
\end{tabular}
\right) $, and $A$ is constant 
\begin{equation}
A=A_0\sqrt{\frac c{-\det M_2}}.  \label{constant}
\end{equation}
Substituting Eq. (\ref{partial3}) into Eq. (\ref{entangled}), we get 
\begin{equation}
E=-A\left[ Z(\beta )\ln A-\frac d{d\beta }Z(\beta )\right] _{\beta =-1},
\label{entanglement}
\end{equation}
in which 
\begin{equation}
Z(\beta )=%
\mathop{\rm tr}
\exp \left\{ -\frac \beta 2(a_1^{\dagger },a_1)N\Sigma _B%
{a_1^{\dagger } \choose a_1}
\right\} .  \label{partition1}
\end{equation}
Using the result in Ref. \cite{pan} 
\begin{eqnarray}
Z(\beta ) &=&%
\mathop{\rm tr}
\exp \left\{ -\frac \beta 2(a_1^{\dagger },a_1)N\Sigma _B%
{a_1^{\dagger } \choose a_1}
\right\}  \nonumber  \label{partition2} \\
\ &=&\left| \det (e^{\beta N}-1)\right| ^{-\frac 12},  \label{partition2}
\end{eqnarray}
direct calculation yields 
\begin{equation}
\frac d{d\beta }Z(\beta )\mid _{\beta =-1}=-\frac 12\left| \det
(e^{-N}-1)\right| ^{-\frac 12}%
\mathop{\rm tr}
\frac N{1-e^N}.  \label{partition3}
\end{equation}
Substituting Eq. (\ref{partition3}) into Eq. (\ref{entanglement}) we get the
explicit expression of the entanglement for the state in Eq. (\ref{density}) 
\begin{equation}
E=-A\left| \det (e^{\beta N}-1)\right| ^{-\frac 12}\left[ \ln A+\frac 12%
\mathop{\rm tr}
\frac N{1-e^N}\right] .  \label{final}
\end{equation}
Thus, from Eq. (\ref{final}) one can conveniently obtain the entanglement of
states in Eq. (\ref{density}) only by calculating the eigenvalues of
negative Hermitian matrix $N$, which can be easily obtained. This is the
main result of this paper. It is worth pointing out that from Eq. (\ref
{final}), the necessary and sufficient condition of separability for the
states in Eq. (\ref{density}) naturally reads 
\begin{equation}
\ln A=\frac 12%
\mathop{\rm tr}
\frac N{e^N-1}.  \label{condition}
\end{equation}

As an application of Eq. (\ref{final}), we consider the entangled states
produced by a beam splitter \cite{Kim} 
\begin{equation}
\left| \psi \right\rangle _{12}=\hat B(\theta ,\phi )\hat S_1(\zeta _1)\hat S%
_2(\zeta _2)\left| 00\right\rangle ,  \label{example1}
\end{equation}
where the beam splitter operator $\hat B(\theta ,\phi )$ is \cite{BS-op} 
\begin{equation}
\hat B(\theta ,\phi )=\exp \left[ \theta (a_1^{\dagger }a_2e^{i\varphi
}-a_1a_2^{\dagger }e^{-i\varphi })\right] ,  \label{operator1}
\end{equation}
and the single-mode squeezed operator $\hat S(\zeta )$ is \cite{squeeze} 
\begin{equation}
\hat S(\zeta )=\exp \left[ \frac 12(\zeta ^{*}a^2-\zeta a^{\dagger
^2})\right] .  \label{operator2}
\end{equation}
Making use of LQTT, we can write the normally ordered form of the density
operator in Eq. (\ref{example1}) as follows 
\begin{eqnarray}
\rho _{12} &=&\left| \psi \right\rangle _{1212}\left\langle \psi \right|
=A_0:\exp \left\{ \frac 12\left[ (a_1^{\dagger },a_1)M_1%
{a_1^{\dagger } \choose a_1}
\right. \right.  \nonumber  \label{density} \\
&&\left. \left. \left. +(a_2^{\dagger },a_2)M_2%
{a_2^{\dagger } \choose a_2}
+2(a_1^{\dagger },a_1)M_{12}%
{a_2^{\dagger } \choose a_2}
\right] \right\} :\right. ,  \label{operator3}
\end{eqnarray}
where 
\begin{eqnarray*}
M_1 &=&-\left( 
\begin{tabular}{ll}
$\alpha $ & $1$ \\ 
$1$ & $\alpha ^{*}$%
\end{tabular}
\right) ,\;M_2=-\left( 
\begin{tabular}{ll}
$\beta $ & $1$ \\ 
$1$ & $\beta ^{*}$%
\end{tabular}
\right) , \\
M_{12} &=&-\left( 
\begin{tabular}{ll}
$\delta $ & $0$ \\ 
$0$ & $\delta ^{*}$%
\end{tabular}
\right) , \\
\alpha &=&\frac{\zeta _1}{\left| \zeta _1\right| }\tanh \left| \zeta
_1\right| \cos ^2\theta +e^{2i\varphi }\frac{\zeta _2}{\left| \zeta
_2\right| }\tanh \left| \zeta _2\right| \sin ^2\theta , \\
\beta &=&e^{-2i\varphi }\frac{\zeta _1}{\left| \zeta _1\right| }\tanh \left|
\zeta _1\right| \sin ^2\theta +\frac{\zeta _2}{\left| \zeta _2\right| }\tanh
\left| \zeta _2\right| \cos ^2\theta , \\
\delta &=&\frac 12\sin \left( 2\theta \right) (\frac{\zeta _2}{\left| \zeta
_2\right| }\tanh \left| \zeta _2\right| e^{i\varphi }-\frac{\zeta _1}{\left|
\zeta _1\right| }\tanh \left| \zeta _1\right| e^{-i\varphi }).
\end{eqnarray*}
Using Eqs. (\ref{constant}) and (\ref{dom}) yields 
\begin{equation}
A=\frac 1{\left| \delta \right| \cosh \left| \zeta _1\cosh \right| \left|
\zeta _2\right| },\;M=\left( 
\begin{tabular}{ll}
$a$ & $d$ \\ 
$b$ & $c$%
\end{tabular}
\right)  \label{result1}
\end{equation}
with 
\begin{eqnarray}
a &=&\frac 1c(1+bd),\;c=\frac{1-\left| \beta \right| ^2}{\left| \delta
\right| ^2},  \nonumber \\
d &=&\frac 1{\left| \delta \right| ^2}\left( \delta ^2\beta ^{*}+\alpha
(1-\left| \beta \right| ^2)\right) ,\;b=-d^{*}.  \label{result2}
\end{eqnarray}
Finally, we get the entanglement of $\left| \psi \right\rangle _{AB}$ in Eq.
(\ref{example1}) 
\begin{eqnarray}
E &=&\frac 1{\left| \delta \right| \cosh \left| \zeta _1\cosh \right| \left|
\zeta _2\right| }  \nonumber  \label{result3} \\
&&\times \frac{\sqrt{\lambda }}{\left| \lambda -1\right| }\left[ \ln (\left|
\delta \right| \cosh \left| \zeta _1\cosh \right| \left| \zeta _2\right|
)\right.  \nonumber  \label{result3} \\
&&\left. -\frac{1+\lambda }{2(1-\lambda )}\ln \lambda \right] ,
\label{result3}
\end{eqnarray}
where $\lambda $ is any of the roots in the following equation 
\begin{equation}
\lambda ^2-(a+c)\lambda +1=0.  \label{result4}
\end{equation}

We can consider a special case of Eq. (\ref{example1}) by choosing $\theta =%
\frac \pi 4$, $\zeta _i=s_ie^{i\varphi }$ $(i=1,2)$ and $\varphi =\frac{l\pi 
}2$ $(l=0,1,2,...)$ \cite{Kim}. In this case Eq. (\ref{result3}) is just the
result of Ref. \cite{Kim}, i.e., 
\begin{equation}
E=\cosh ^2\left| s\right| \ln (\cosh ^2\left| s\right| )-\sinh ^2\left|
s\right| \ln (\sinh ^2\left| s\right| ).  \label{special1}
\end{equation}
where $s=\frac 12(s_1e^{i\varphi }+s_2e^{-i\varphi })$. Furthermore, if we
choose $\varphi =0$, $\zeta _2=-\zeta _1=r$ and $\theta =\frac \pi 4$, we
get $\lambda _1=\tanh ^2r$. we then get the well known result for the
two-mode squeezed vacuum state, whose entanglement is \cite{Parker} 
\begin{equation}
E=\cosh ^2r\ln (\cosh ^2r)-\sinh ^2r\ln (\sinh ^2r).  \label{special2}
\end{equation}

For non-Gaussian states, generally speaking, their entanglement can not be
written in the compact form as in Eq. (\ref{final}). But for some states
such as $SU(2)$ coherent states (see, e.g., \cite{Kim}) and the photon-added
coherent states \cite{Agarwal}, the entanglement can be easily obtained by
solving the eigenvalue problem of the reduced operator of Eq. (\ref
{addition3}). To show this, we consider the following example \cite{Kim} 
\begin{equation}
\left| \psi \right\rangle =\hat B\left| n_1,n_2\right\rangle =e^{\theta
(a_1^{\dagger }a_2-a_2^{\dagger }a_1)}\left| n_1,n_2\right\rangle
\label{example2}
\end{equation}
The density operator $\rho _{12\text{ }}$of this system can be written, by
following the above procedure, as 
\begin{eqnarray}
\rho _{12\text{ }} &=&\hat B\left| n_1,n_2\right\rangle \left\langle
n_2,n_1\right| \hat B^{\dagger }=\frac 1{n_{1!}n_2!}\frac{d^{n_1}}{d\alpha
^{n_1}}\frac{d^{n_2}}{d\beta ^{n_2}}  \nonumber \\
\times &:&\exp \left\{ \frac 12\left[ (a_1^{\dagger },a_1)M_1%
{a_1^{\dagger } \choose a_1}
+(a_2^{\dagger },a_2)M_2%
{a_2^{\dagger } \choose a_2}
\right. \right.  \nonumber  \label{non-Gauss1} \\
&&\ \left. \left. \left. +2(a_1^{\dagger },a_1)M_{12}%
{a_2^{\dagger } \choose a_2}
\right] \right\} :\mid _{\alpha =\beta =0}\right. .  \label{non-Gauss1}
\end{eqnarray}
Here 
\begin{eqnarray}
M_1 &=&\left( \alpha \cos ^2\theta +\beta \sin ^2\theta -1\right) \sigma _1,
\nonumber \\
M_2 &=&\left( \beta \cos ^2\theta +\alpha \sin ^2\theta -1\right) \sigma _1,
\nonumber \\
M_3 &=&\sin \theta \cos \theta (\beta -\alpha )\sigma _1,  \label{non-Gauss2}
\end{eqnarray}
where $\sigma _1$ is Pauli matrix. Using Eq. (\ref{addition3}), we derive 
\begin{eqnarray}
\rho _A &=&\frac 1{n_{1!}n_2!}\frac{d^{n_1}}{d\alpha ^{n_1}}\frac{d^{n_2}}{%
d\beta ^{n_2}}\frac 1{\left( \beta \cos ^2\theta +\alpha \sin ^2\theta
-1\right) }  \nonumber \\
&&\ \left. \times \left( \frac{\alpha \beta -(\alpha \cos ^2\theta +\beta
\sin ^2\theta )}{\left( \beta \cos ^2\theta +\alpha \sin ^2\theta -1\right) }%
\right) ^{a_1^{\dagger }a_1}\mid _{\alpha =\beta =0}\right. .
\label{non-Gauss3}
\end{eqnarray}
Thus we get the eigenvalues as follows 
\begin{eqnarray}
\lambda _{N_1} &=&\frac 1{n_{1!}n_2!}\frac{d^{n_1}}{d\alpha ^{n_1}}\frac{%
d^{n_2}}{d\beta ^{n_2}}\frac 1{\beta \cos ^2\theta +\alpha \sin ^2\theta } 
\nonumber  \label{non-Gauss4} \\
&&\left. \times \left[ 1+\frac{\alpha \beta -1}{\beta \cos ^2\theta +\alpha
\sin ^2\theta }\right] ^{N_1}\mid _{\alpha =\beta =-1}\right.
\label{non-Gauss4}
\end{eqnarray}
Note that in Eq. (\ref{non-Gauss4}) we take the replacements $\alpha =\alpha
^{^{\prime }}+1$ and $\beta =\beta ^{^{\prime }}+1$. Finally, we get the
entanglement of the state in Eq. (\ref{example2}) 
\begin{equation}
E=-\sum_{N_1}\lambda _{N_1}\ln \lambda _{N_1}.  \label{final2}
\end{equation}

In summary, we have presented a general formula of reduced density operator
for any biparticle pure entangled state of continuous variables. Based on
this formula, we have derived the explicit expression of the entanglement
for any Gaussian pure state in a compact form. From the examples mentioned
above, it is easy to see that Gaussian states' entanglement can be
conveniently calculated by the formula given in this paper. For the case of
non-Gaussian states, we have also shown the usefulness of the formula for
calculating the reduced density operator, as given above. In this sense, the
method to calculate the entanglement for biparticle pure entangled state is
easier to operate.

This work was supported by the National Natural Science Foundation of China
under Grants No. 19975043, No. 10104014 and No. 10028406, and by the Chinese
Academy of Sciences.

\end{document}